\documentclass[11pt]{article}
\usepackage[a4paper,hmarginratio=1:1,vmarginratio=2:3,totalwidth=15.8cm,
totalheight=21.5cm]{geometry}
\usepackage{bm,epstopdf,epsfig,amsmath,amssymb,amsfonts,
latexsym,cancel,verbatim,ulem,color}
\usepackage{amsfonts}
\usepackage{amssymb}

\usepackage{graphicx}
\newcommand{\revision}[1]{\textcolor{black}{#1}}
\newcommand{\strike}[1]{}
\newcommand{\newrevision}[1]{\textcolor{black}{#1}}

\long
\def\symbolfootnote[#1]#2{\begingroup
\def\thefootnote{\fnsymbol{footnote}}
\footnote[#1]{#2}\endgroup} 
\providecommand{\U}[1]{\protect\rule{.1in}{.1in}}

\def\beq{\begin{equation}}
\def\eeq{\end{equation}}
\def\bea{\begin{eqnarray}}
\def\eea{\end{eqnarray}}
\def\bq{\begin{quote}}
\def\eq{\end{quote}}

\def\tu{\tilde u}

\def\tst{\tilde t}
\def\ttau{\tilde \tau}

\def\tg{\tilde g}

\def\tq{\tilde q}

\def\tw{\tilde{\chi}^+}
\def\tz{\tilde{\chi}^0}

\def\bq{\begin{quote}}
\def\eq{\end{quote}}

\renewcommand{\baselinestretch}{1.25}
\begin{document}
%%%%%%%%%%%%%%%%%%%%%%%%%%%%%%%%%%%%%%%%%%%%%%%%%%%%%%%%%%%%%%%%%%%%%%%%%%%%%%%

\thispagestyle{empty}

\vspace*{-2.4cm}
\begin{flushright}
  CERN-PH-TH/2010-005\\
  CPHT-RR001.0111\\
  LPSC-11026\\
  OUTP-11-32P
\end{flushright}

\vspace*{4mm}

\begin{center}
{\Large \textbf{Fine-tuning implications for complementary dark matter and\\[2mm] LHC SUSY searches}}

\vspace*{8mm}

\textbf{S.~Cassel$^{\,\,a}$, 
D.~M. Ghilencea$^{\,\,b,c,}\!\!$\symbolfootnote[4]{on leave from Theoretical Physics Department, IFIN-HH Bucharest MG-6, Romania}, 
S.~Kraml$^{\,\,d}$,
A.~Lessa$^{\,\,e\,}$,
G.~G.~Ross$^{\,\,a\,}$\symbolfootnote[2]{e-mail~addresses:
s.cassel1@physics.ox.ac.uk, dumitru.ghilencea@cern.ch, sabine.kraml@lpsc.in2p3.fr,\\ 
lessa@nhn.ou.edu, g.ross1@physics.ox.ac.uk}} \\[0pt]

\vspace{0.5cm} {$^a\, $ Rudolf\, Peierls\, Centre for Theoretical Physics,
\,University\, of\, Oxford,\\[0pt]1 Keble Road, Oxford OX1 3NP, United Kingdom.}
\\[3pt]

{\ $^b \, $ Department of Physics, CERN - Theory Division, CH-1211 Geneva 23,
Switzerland.}\\[3pt]

{\ $^c \, $ Centre de Physique Theorique, Ecole Polytechnique, CNRS, 91128
Palaiseau, France.}\\[1pt]

{\ $^d\, $ Laboratoire de Physique Subatomique et de Cosmologie, UJF Grenoble 1,
CNRS/IN2P3, INPG, 53 Avenue des Martyrs, F-38026 Grenoble, France.}\\[3pt]

{\ $^e\, $ Dept. of Physics and Astronomy, University of Oklahoma, Norman, OK 73019, USA.}%\\[6pt]

\end{center}

\bigskip

\begin{abstract}\noindent
The requirement that SUSY should solve the hierarchy problem without undue 
fine-tuning imposes severe constraints on the new supersymmetric states. 
With the MSSM spectrum and soft SUSY breaking originating from universal 
scalar and gaugino masses at the Grand Unification scale, we show that 
the low-fine-tuned regions fall into two classes that will require 
complementary collider and dark matter searches to explore in the 
near future. The first class has relatively light gluinos or squarks 
which should be found by the LHC in its first run. We identify the 
multijet plus $E_{T}^{miss}$ signal as the optimal channel 
%for the LHC gluino discovery 
and determine the discovery potential in the first run. 
The second class has heavier gluinos and squarks but the LSP has a 
significant Higgsino component and should be seen by the next 
generation of direct dark matter detection experiments. 
The combined information from the 7~TeV LHC run and the next generation 
of direct detection experiments can test almost all of the CMSSM parameter 
space consistent with dark matter and EW constraints, corresponding to 
a fine-tuning not worse than 1:100. To cover the complete low-fine-tuned 
region by SUSY searches at the LHC will require running at the full 
14~TeV CM energy;  in addition it may be tested indirectly by Higgs 
searches covering the mass range below 120~GeV.
\end{abstract}

\bigskip

\clearpage
%---------------------------------------------------------------------------------
\section{Introduction}
%---------------------------------------------------------------------------------

Weak scale supersymmetry (SUSY) has been proposed as a solution to the 
hierarchy problem, {\it i.e.}\ it ensures that the electroweak breaking 
scale is consistent with radiative corrections without undue fine-tuning. 
However, to achieve this, the new SUSY states must be relatively light. 
To quantify how light the SUSY states should be and 
and how much stress experimental limits already put on SUSY, one can apply  
a measure of fine-tuning. 
In~\cite{Cassel} an analysis using SOFTSUSY~\cite{Allanach:2001kg} 
was made of the status of SUSY searches in the constrained minimal 
supersymmetric standard model (CMSSM) using the electroweak (EW) 
fine-tuning measure, $\Delta$, introduced in~\cite{Ellis:1986yg} 
computed to two-loop order. 
A scan of the CMSSM parameter space was performed, requiring acceptable 
radiative electroweak breaking, non-tachyonic SUSY particle masses 
(avoiding colour and charge breaking vacua), consistency with the 
experimental bounds on superpartner 
masses and with constraints from BR$(b\rightarrow s\gamma)$, 
BR$(B_{s}\rightarrow\mu^{+}\mu^{-})$ and the muon $(g-2)$ 
as detailed in~\cite{Cassel}.

\begin{figure}[t]
\centering
\includegraphics[width=9cm]{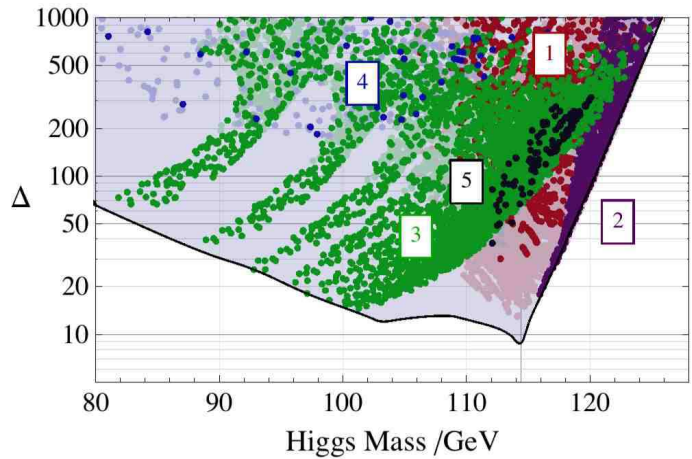} 
\def\baselinestretch{1.}
\caption{{\protect\small Two-loop fine-tuning versus Higgs mass 
for the scan over CMSSM parameters with no constraint on the 
Higgs mass. The solid line is the minimum fine-tuning with 
$\left( \protect\alpha_s^{}, M_t^{} \right) =$ (0.1176, 173.1~GeV). 
The dark green, purple, crimson and black coloured regions have a 
dark matter density within
$\Omega h^{2} = 0.1099 \pm 3 \times 0.0062$~\cite{Dunkley:2008ie}
(i.e. 3$\sigma$ saturation)
while the lighter coloured versions of these regions lie below this 
bound. The colours and their associated numbers refer to different LSP 
structures as described in the text. Regions 1, 3, 4 and 5 have 
an LSP which is mostly bino-like. In region~2, the LSP has a 
significant higgsino component. \newrevision{Representative phase space  points for 
regions 1,2...5,  denoted SUG1, SUG2,...,SUG5,  respectively, will be 
analysed in detail later on,  together with the  point of minimal EW 
$\Delta\approx 9$  denoted SUG0.} }}
\label{higgsp}
\end{figure}

In Figure \ref{higgsp} the envelope of the shaded region shows the 
EW fine-tuning \newrevision{$\Delta$}. One may see that, imposing all the constraints 
listed above except for the LEPII bound on the Higgs mass, there is a 
minimum of \newrevision{the EW} fine-tuning, $\Delta\approx 9$, for a Higgs mass of 
$m_h\approx 114$~GeV.\footnote{Note that the calculation of $m_h$ 
is subject to a theoretical uncertainty of about 
2--3~GeV~\cite{Degrassi:2002fi}, see also \cite{Cassel}.} 
\strike{Although we consider here only the low-fine-tuned regions of the CMSSM, this 
covers a large part of the low-fine-tuned MSSM parameter space because many 
models beyond the CMSSM are more fine-tuned. For example gauge mediation is more 
fine-tuned because the former relies on the scalar focus point~\cite{Feng:2000bp}  
whereas the latter does not have the focus point as it does not have degenerate 
scalars at a high scale.}
\revision{Although the analysis presented here is concerned with the CMSSM, this class of model permits the presence of a scalar focus point, which favours small fine tuning of the electroweak scale~\cite{Feng:2000bp} relative to more generic MSSM models. %Models without gravity mediation SUSY breaking tend to require increased fine tuning.
}
Thus if one excludes the low-fine-tuned regions of 
the CMSSM one can say that much larger ranges of the MSSM parameter space 
are also disfavoured. %excluded. 
To date the only class of non-CMSSM models with MSSM 
spectrum that have lower fine-tuning are those with non-universal gaugino 
masses with a `natural' relation between them that reduces the gluino 
mass relative to the CMSSM case \revision{ \cite{Gogoladze:2009bd,Horton:2009ed} }. 
 
In addition to EW fine-tuning, another important constraint is that the 
thermal relic abundance of the lightest supersymmetric particle (LSP), 
which contributes to dark matter, should be consistent with cosmological 
observations, \revision{under the assumption of R-parity conservation}. 
Here we use MicrOMEGAs~\cite{Belanger:2006is} as computational tool.  
Imposing the constraints of \cite{Dunkley:2008ie}  
limits the allowed region to be in the coloured regions of Figure~\ref{higgsp}. 
In regions~1 (red) and 5 (black points superposed on the green region) $h^{0}$ 
and $A^{0}/H^{0}$ resonant annihilation respectively are responsible for 
reducing the dark matter abundance within current bounds.  
Region~2 (purple) has significant bino-higgsino mixing in the LSP, 
and annihilation proceeds via higgsino t-channel exchange to EW gauge bosons. 
Finally, for regions~3 (green) and 4 (blue), the dominant processes are stau 
and stop co-annihilation, respectively. Overall, requiring that the SUSY 
dark matter relic density should be within present bounds raises the 
minimal fine-tuning to $\Delta=15$, still quite reasonable, corresponding 
to $m_{h}\approx 115$~GeV. % $m_{h}=114.7\pm 2$~GeV.

%%% Fig 1 %%%

An immediate question is what is the best way to test the low-fine-tuned 
region of SUSY parameter space and will it be tested soon? In this the 
search for the Higgs boson plays a very important role because, 
{\it c.f.} Figure~\ref{higgsp}, for large Higgs masses, quantum corrections 
make the fine-tuning exponentially sensitive to the Higgs mass. Thus, 
if the Higgs is not found below $\approx$120~GeV, it will imply that 
the fine-tuning is uncomfortably %unacceptably 
large, $\Delta>100$.  
However, a LHC Higgs discovery at such low mass will be difficult 
and is likely to take several years. 
Given this, it is of interest to consider to what extent direct SUSY 
searches will probe the regions of low fine-tuning.

In this paper we consider both collider searches and dark  matter searches. 
For this it is important to analyse the nature of the SUSY spectrum in the 
region of low fine-tuning ($\Delta<100$) in light of the Higgs mass and dark 
matter constraints.  
Of course, if the LHC is to detect new SUSY states in its first run with 
of ${\cal{O}}(1)$fb$^{-1}$ luminosity at a CM energy of 7~TeV, some states 
with coupling to the gluon, {\it i.e.}\ squarks or gluinos, must be light 
enough to have a sizeable cross section. 
In \cite{Baer:2010tk}, the discovery reach of the LHC %in its first run 
with 1~(2)~fb$^{-1}$ luminosity at a CM energy of 7~TeV  
was determined as 
$m_{\tg}\sim 1100~(1200)$~GeV for $m_{\tq}\sim m_{\tg}$, and 
$m_{\tg}\sim 620~(700)$~GeV for $m_{\tq}\gg m_{\tg}$.  
The first results from CMS \cite{cms} for 35~pb$^{-1}$ of data  
exclude gluino masses below 500~GeV for $m_0\lesssim 350$~GeV, 
but no limit is obtained for $m_0\gtrsim 500$~GeV. 

%%% Fig 2 %%%
\begin{figure}[t]
\centering\def\baselinestretch{1.}
\begin{tabular}{cc} 
  \includegraphics[width=0.48\textwidth]{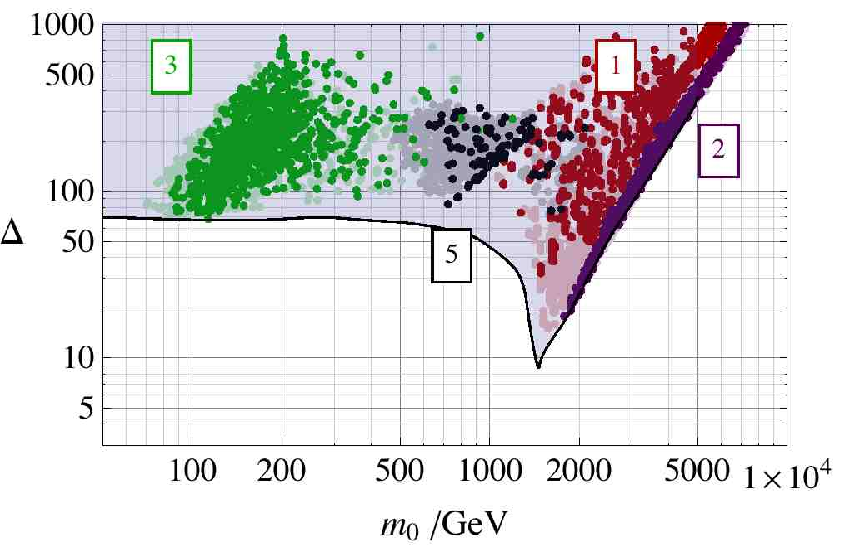}&
  \includegraphics[width=0.48\textwidth]{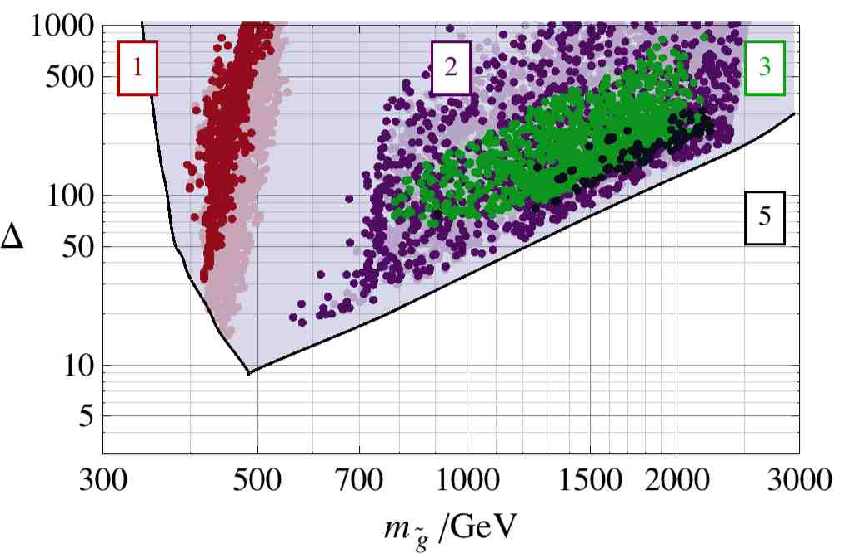}\\
  (a) & (b) 
\end{tabular}
\caption{
{\protect\small (a) Fine-tuning versus scalar mass parameter,  
(b) fine-tuning versus gluino mass; 
in both cases the constraint on the Higgs mass, $m_{h}>114.4$~GeV is applied.}
}
\label{higgsp1}
\end{figure}
 
\begin{figure}[t]
\centering\def\baselinestretch{1.}
\begin{tabular}{cc} 
  \includegraphics[width=0.48\textwidth]{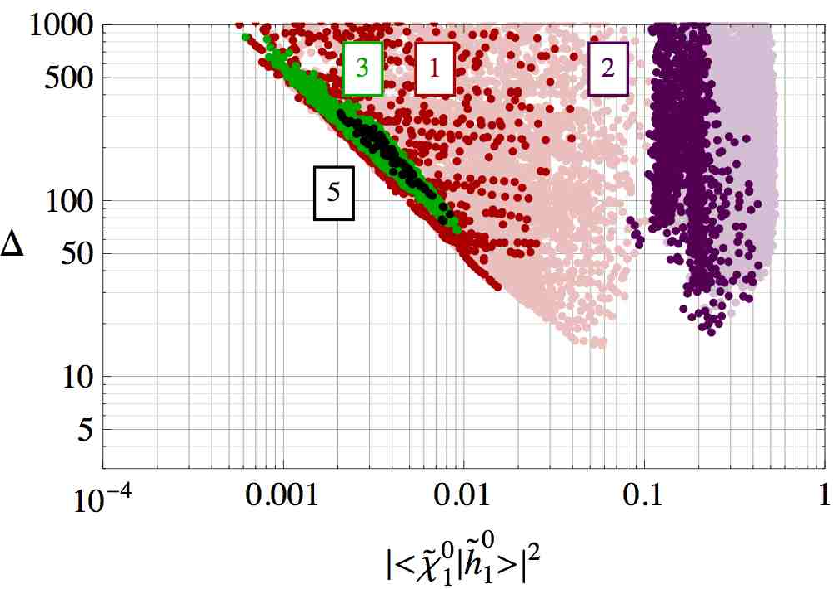}&
  \includegraphics[width=0.48\textwidth]{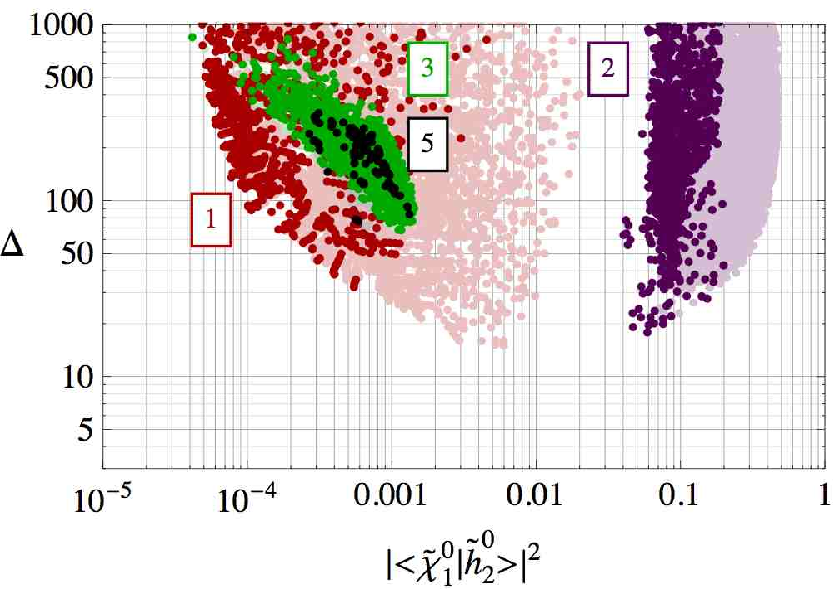}\\
  (a) LSP $\tilde{h}_{1}^{0}$ component & (b) LSP $\tilde{h}_{2}^{0}$ component
\end{tabular}
\caption{{\protect\small The higgsino components in the LSP, again requiring 
$m_{h}>114.4$~GeV.}}
\label{higgsp2}
\end{figure}

\begin{figure}[t]\centering
\def\baselinestretch{1.}
\begin{tabular}{cc} 
  \includegraphics[width=0.48\textwidth]{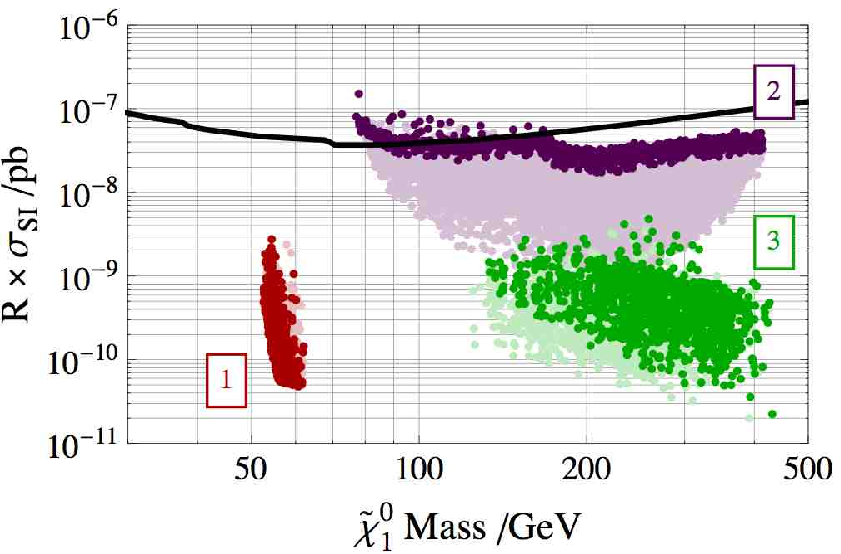} & 
  \includegraphics[width=0.48\textwidth]{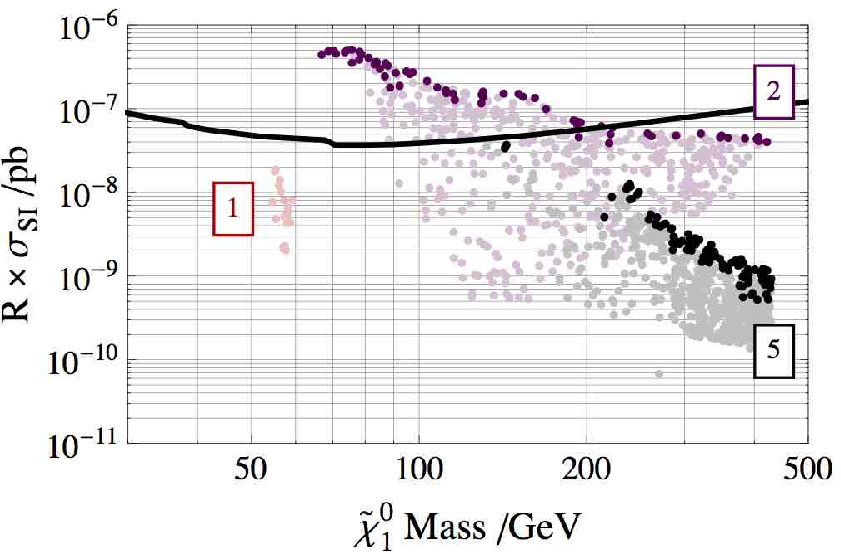}\\
  (a) $\tan \beta \leq 45$ & (b) $50 \leq \tan \beta \leq 55$
\end{tabular}
\caption{
{\protect\small Scaled spin independent cross section for 
LSP-proton scattering, with $m_{h}>114.4$~GeV. The scaling factor 
$R=\Omega h^{2}/0.1099$ has been applied. The solid line is the CDMS-II limit. 
All points satisfy $\Omega h^{2} < 0.1285$, with those with darker 
shading lying within $3\sigma$ of the WMAP bound, 
$\Omega h^{2} = 0.1099 \pm 3 \times 0.0062$.}
\revision{The mSUGRA phase space scan was discrete only in the $\tan \beta$ dimension with 30 slices in the range $2 \leq \tan \beta \leq 45$, and two further slices at $\tan \beta = 50, 55$.}
}
\label{dmcdmsbound}
\end{figure}

How does this compare to the expectation for the regions of low fine-tuning? 
From Figure~\ref{higgsp1}(a) one sees that these regions %of low fine-tuning 
(which we take as $\Delta<100$) can either have light squarks (region~3, 
green points) 
or be close to the scalar ``focus point'' corresponding to heavy 
squarks.\footnote{The few points of region 4 that 
satisfiy the Higgs mass limit have very large fine-tuning, $\Delta\sim 10^3$, 
and are not shown in Figures~\ref{higgsp1}--\ref{dmcdmsbound}. 
\revision{Relaxing the Higgs mass constraint by 2-3~GeV has a negligible effect on the plots, except for region 3 where reduced fine tuning ($\Delta \gtrsim 45$) becomes possible}} 
Figure~\ref{higgsp1}(b) shows that the low fine-tuned points with heavy 
squarks (regions 1, 2 and 5) have two components. The first, region 1, 
has a small gaugino mass parameter and corresponds to gluinos with mass 
of about 400--500~GeV, 
potentially accessible to LHC discovery in the first run. 
The remaining regions (2 and 5) have gluino masses 
beyond the LHC reach in the 7~TeV run. 
However these regions may be accessible to dark matter searches. 
Such searches put limits on, {\it e.g.}, the spin-independent 
scattering cross section 
for neutralino dark matter and this in turn is dominated by Higgs-boson 
exchange coupling to the higgsino and bino components in the neutralino. 
Typically the LSP has a large bino component but in restricted regions 
of parameter space it may also have a sizeable higgsino component. 
In Figure~\ref{higgsp2} we plot the higgsino component versus the 
fine-tuning measure. One can see that only the purple points, region~2, 
have a significant higgsino component. 

%%% Fig 3 %%%

\begin{figure}[t]\centering
\def\baselinestretch{1.}
\begin{tabular}{cc} 
  \includegraphics[width=0.48\textwidth]{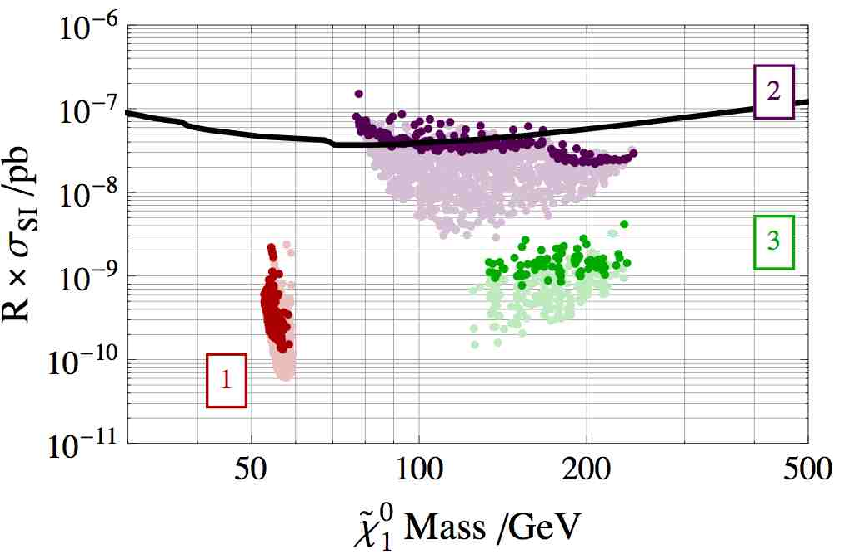} & 
  \includegraphics[width=0.48\textwidth]{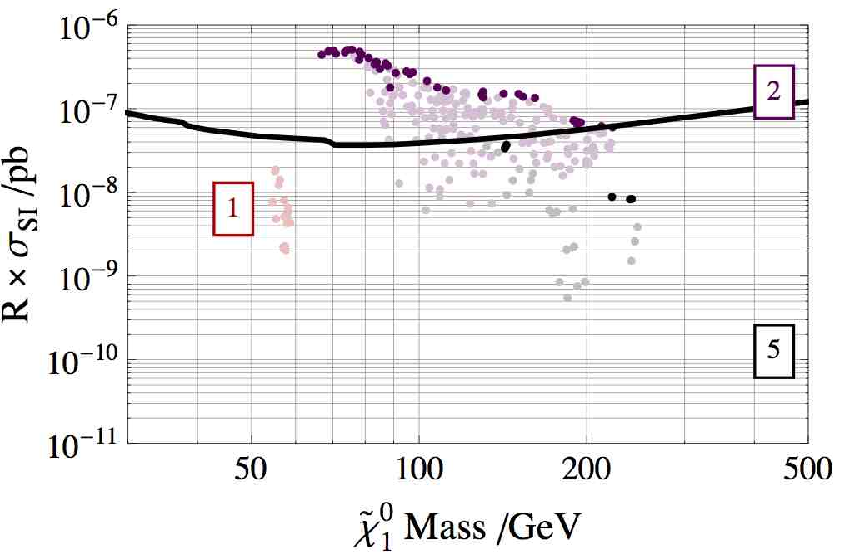}\\
  (a) $\tan \beta \leq 45$ & (b) $50 \leq \tan \beta \leq 55$
\end{tabular}
\def\baselinestretch{1.}
\caption{
{\protect\small Same as Figure~\ref{dmcdmsbound} but imposing in addition $\Delta<100$. }
}
\label{dmcdmsboundft100}
\end{figure}

Once one knows the composition of the LSP it is straightforward to compute 
the spin independent cross section relevant to the direct dark matter searches. 
This is shown in Figure~\ref{dmcdmsbound}, plotted against the LSP mass. 
Here, the cross-section has been rescaled by $R= \Omega h^{2}/0.1099$, 
to take into account of the dark matter abundance at each of the points. 
Also shown is the current best bound coming from the CDMS 
experiment~\cite{Ahmed:2009zw}. For more details see \cite{cassel-thesis}. 
One sees that this already provides a significant test of region 2. 
A factor of 10 improvement in the dark matter sensitivity, which should 
be achieved by SuperCDMS in 2013, will probe almost the full range of 
region 2, the exception being points that do not saturate the dark matter 
density. For these latter points a two orders of magnitude improvement 
will be needed. The points in region 1, 3 and 5 have very small higgsino 
component and dark matter searches do not test this region. However, as 
may be seen from Figure~\ref{higgsp}, the bulk of this data is very 
fine-tuned with $\Delta>100$. 
The situation for $\Delta<100$ is depicted in Figure~\ref{dmcdmsboundft100}. 
We see that the parameter space still needing to be explored has shrunk 
considerably: only a small part of regions 1 and 3 and an even smaller part 
of region 5 remains to be tested.

\begin{figure}[t]
\centering\def\baselinestretch{1.}
\begin{tabular}{cc} 
  \includegraphics[width=0.48\textwidth]{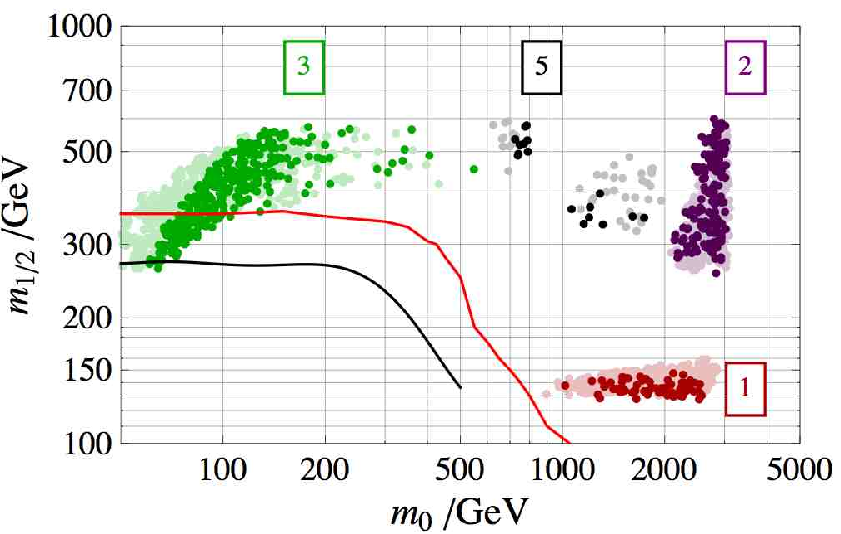} & 
  \includegraphics[width=0.48\textwidth]{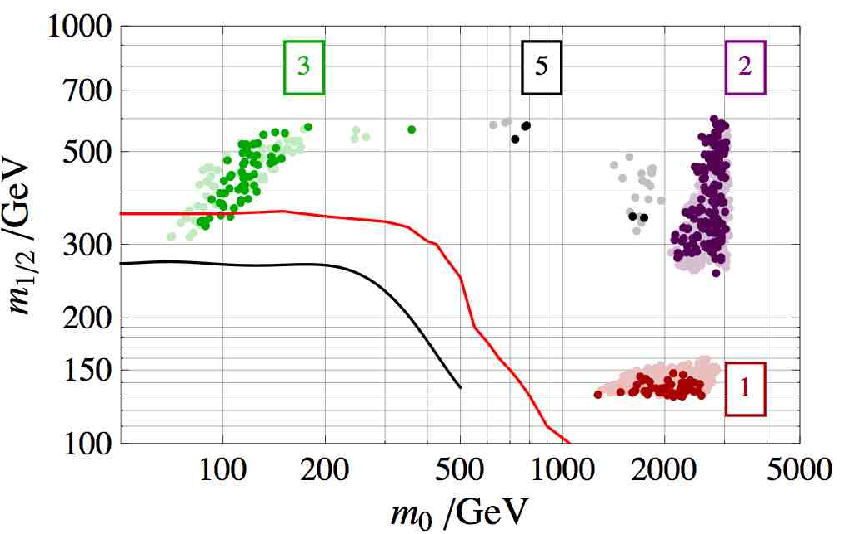}\\
  (a) $m_h^{} > 111$\,GeV & (b) $m_h^{} > 114.4$\,GeV
\end{tabular}
\caption{{\protect\small Regions of low fine-tuning ($\Delta <100$) 
in the $m_0$ versus $m_{1/2}$ plane, summed over $\tan\beta$ and $A_0$.  
All points satisfy the SUSY and Higgs mass limits, $\Omega h^{2} < 0.1285$ 
(dark points having $0.0913<\Omega h^{2} < 0.1285$), 
the B-physics and $\delta a_\mu$ constraints, and the 
CDMS-II bound on the dark matter detection cross section. 
The area below the \strike{red} \revision{black} line shows the CMSSM exclusion 
(for $\tan\beta=3$ and $A_0=0$) from the CMS dijet$+E_T^{miss}$ 
analysis \cite{cms},
\revision{and that below the red line the ATLAS exclusion area \cite{daCosta:2011qk}}. 
\newrevision{
The different lower bounds on $m_h$ of these plots are applied to show the impact of the
2-3 GeV theoretical uncertainty in the calculation of $m_h$ \cite{Cassel,Degrassi:2002fi}.}}}
\label{m0m12}
\end{figure}

These results  are also  illustrated in Figure~\ref{m0m12} in the 
$(m_0,m_{1/2})$ plane,  which shows the same points of low fine-tuning that
pass  the constraints mentioned. Notice that all these points are also
consistent with the latest  CMS observed exclusion area~\cite{cms} 
(situated below the \strike{red} \revision{black} curve), \revision{and all but a fraction of the region 3 points are consistent with the latest ATLAS exclusion limit \cite{daCosta:2011qk} }. 
We will discuss below the LHC configuration needed to scan the SUSY spectrum 
for these residual regions. 

We conclude this Introduction with a side-remark: 
one often states that the very small area of points left
in the moduli space $(m_0, m_{1/2})$, that respect all
experimental  constraints, renders supersymmetry an unlikely solution
to the hierarchy problem.  However, even if this area is reduced to
few points due to further experimental constraints,  recall  that many
of  them have acceptable  fine-tuning (in our case $\Delta<100$).  
That is, the density and size of the area of points  allowed
in Figure~\ref{m0m12} does not necessarily have a physical relevance, 
and  cannot be used to conclude that only few points left would immediately
invalidate supersymmetry  as a solution to the hierarchy problem. It
would rather indicate the most likely values of these moduli $(m_0,m_{1/2})$,  
that a fundamental  theory beyond MSSM should fix dynamically, to avoid 
degenerate  vacua. 

%%% Fig 4 %%% 

%%% Fig 5 %%% 

%%% Fig 6 %%% 

%---------------------------------------------------------------------------------
\section{Testing the low fine-tuned regions at the LHC}
%---------------------------------------------------------------------------------

We turn now to a discussion of the LHC sensitivity to the low-fine-tuned SUSY 
regions of parameter space. The LHC searches are complementary to the dark matter 
searches, being sensitive to the %low fine-tuned 
points with a light gluino or squark, which correspond to a LSP with a spin 
independent cross section that is too small to be seen by the dark matter searches. 
Therefore it is of interest to consider the prospects for the LHC to probe the 
remaining regions of parameter space that will not be probed by the dark matter 
searches even though they may have heavy coloured states beyond the reach  
of the 7~TeV run. 
Accordingly, we consider five representative points, 
listed in Table~\ref{tab:bm}, \newrevision{see also Figure 1}, which we will study in detail:
SUG0 (point of minimal fine-tuning $\Delta\sim 9$), 
SUG1 (of region 1, red), SUG2 (of region 2, purple), 
SUG3 (of region 3, green), SUG5 (of region 5, black).
\newrevision{Unlike the SUG0 point, the  points SUG1, SUG2,..., SUG5 
have neutralino   dark matter within $3\sigma$ of the observed dark
matter abundance.} 
 
\begin{table}\centering
\begin{tabular}{lrrrrr}
\hline
 & SUG0  & SUG1 & SUG2 & SUG3 & SUG5\\
\hline
$m_0$   & 1455 & 1508 & 2270 & 113 & 725\\
$m_{1/2}$ & 160 & 135 & 329 & 383 & 535\\
$A_0$ & 238 & 1492 & 30 & -220 & 1138\\
$\tan\beta$ & 22.5 & 22.5 & 35 & 15 & 50\\
\hline
$\mu$  & 191 & 433 & 187 & 529 & 581 \\
$m_{\tg}$ & 482 & 414 & 900 & 898 & 1252\\
$m_{\tu_L}$ & 1469 & 1509 & 2331  & 826 & 1315\\
$m_{\tst_1}$ & 876 & 831 & 1423 & 602 & 1000\\
\hline
$m_{\tw_1}$ & 106 & 104 & 168 & 293 & 416\\
$m_{\tz_2}$ & 108 & 104 & 181 & 293 & 416\\ 
$m_{\tz_1}$ & 60 & 53 & 123  & 155 & 222\\ 
\hline
$\Delta$ & 9 & 50 & 45 & 68 & 84 \\
$\Omega_{\tz_1} h^2$ & 0.41 & 0.13 & 0.10 & 0.13 & 0.10\\
\hline
BR$(b\rightarrow s\gamma) \times 10^{4}$ & 3.4 & 3.7 & 3.4 & 3.2 & 3.2\\
BR$(B_{s} \rightarrow \mu^{+}\mu^{-})\times 10^{9} $ & 3.0 & 2.9 & 2.9 & 3.4 & 1.7\\
$\delta a_{\mu} \times 10^{10}$ & 4.5 & 3.2 & 3.2 & 22.5 & 16.6\\
%\hline
$\sigma^{\rm SI}_{\chi p}$ (pb) $\times 10^{10}$ & 108 & 5 & 432 & 24 & 101 \\
\hline
$\sigma^{(LO)}(7$ TeV) (pb)  & 8 & 12 & 0.9 & 0.4 & 0.02 \\
$\sigma^{(LO)}(14$ TeV) (pb) & 40 & 75 & 3 & 5 & 0.4  \\
\hline
\end{tabular}\\[2mm]
\caption{CMSSM parameters and sparticle masses in~GeV for the points used in 
our LHC analysis.
We also show for each of the points 
the amount of fine-tuning, the neutralino relic density, 
the branching ratios of $b\rightarrow s\gamma$ and $B_s\to\mu^{+}\mu^{-}$, 
the SUSY contribution to the muon anomalous magnetic moment $\delta a_{\mu}$, 
the spin-independent LSP scattering cross section off protons 
$\sigma^{\rm SI}_{\chi p}$, and the total leading-order sparticle production 
cross-sections for the LHC at $\sqrt{s} = 7$ and 14~TeV.}
\label{tab:bm}
\end{table}

The first point, SUG0, has a gluino mass of 482 GeV and a total cross-section 
of 8~(40)~pb at $\sqrt{s} = 7$ (14) TeV. Although this point has a too  
large neutralino relic abundance, \revision{it possesses the smallest fine 
tuning found when imposing the non-dark-matter experimental constraints. 
We include it here for the purpose of comparison 
\newrevision{with the other representative points. 
The  SUG0 point could also respect the dark matter constraint}} 
% It could be modified in a way that changes the dark matter constraints 
without changing the EW fine-tuning significantly, for example by adding 
a small amount of R-parity violation causing the LSP to 
decay~\cite{Buchmuller:2007ui}, or if the true dark matter consists 
of  axions/axinos~\cite{Baer:2010gr,Covi2}. 
\strike{
Moreover, the gluino phenomenology of this point is quite similar to that 
of a nearby light red point in Figure~\ref{higgsp} that has a small 
neutralino abundance. }
\newrevision{Indeed, as shown in \cite{Covi2} for an axino LSP the usual
dark matter constraint can  be relaxed. In this case
the SUG0 point could also satisfy the relic abundance constraint, 
without changing its LHC phenomenology, due to the smallness of 
the axino coupling.}

% As mentioned, the remaining points have neutralino  
% dark matter within $3\sigma$ of the observed dark matter abundance. 
The second representative point (SUG1) corresponds to a dark red point 
in Figure~\ref{higgsp} with moderate fine-tuning, $\Delta \sim 30$.  
As may be seen from Figure~\ref{higgsp1}(b), the dark red points all have 
low gluino mass; 
for the point chosen $m_{\tilde g}=414$ GeV and the total 
cross-section is 12 (75) pb at $\sqrt{s} = 7$ (14) TeV. The LSP 
is predominantly bino with a scattering cross section off nuclei
that is too small to be probed by the next generation of direct dark 
matter searches.

The third representative point (SUG2) is chosen to lie in region 2, close to 
the CDMS bound. It has a heavy gluino, $m_{\tilde g}=900$~GeV, and TeV-scale 
squarks, resulting in a total cross-section of 0.9 (3) pb at 
$\sqrt{s} = 7$ (14) TeV, mostly dominated by chargino-pair production. 
The LSP has significant bino and higgsino components and therefore a 
sizable $\sigma^{\rm SI}_{\chi p}$.

The fourth representative point (SUG3) lies in region 3 and has the lowest 
fine-tuning, $\Delta=68$, in that region. It saturates the dark matter 
density and might be probed by direct dark matter searches if the sensitivities 
can be improved by more than an order of magnitude. It has a gluino mass of 
898~GeV and an LSP with a mass of 155~GeV that is almost a pure bino. 
The total sparticle production cross-section at $\sqrt{s} = 7$ (14) TeV 
is 0.4 (5) pb.

The fifth and final representative point (SUG5) is in region 5 and also has 
the lowest fine-tuning, $\Delta=84$, in that region. It saturates the dark 
matter density and may be probed by direct dark matter searches in the 
near future. It has both gluinos and squarks at the TeV scale, with a 
sparticle production cross-section of only 0.02 (0.4) pb at $\sqrt{s} = 7$ 
(14) TeV. The LSP has a mass of 222~GeV and is almost a pure bino.

%-----------------------------------------------------------------------------
\subsection{LHC at 7 TeV and 1 fb$^{-1}$}
%-----------------------------------------------------------------------------

\begin{figure}[t]
\centering\def\baselinestretch{1.}
(a)\includegraphics[width=0.46\textwidth]{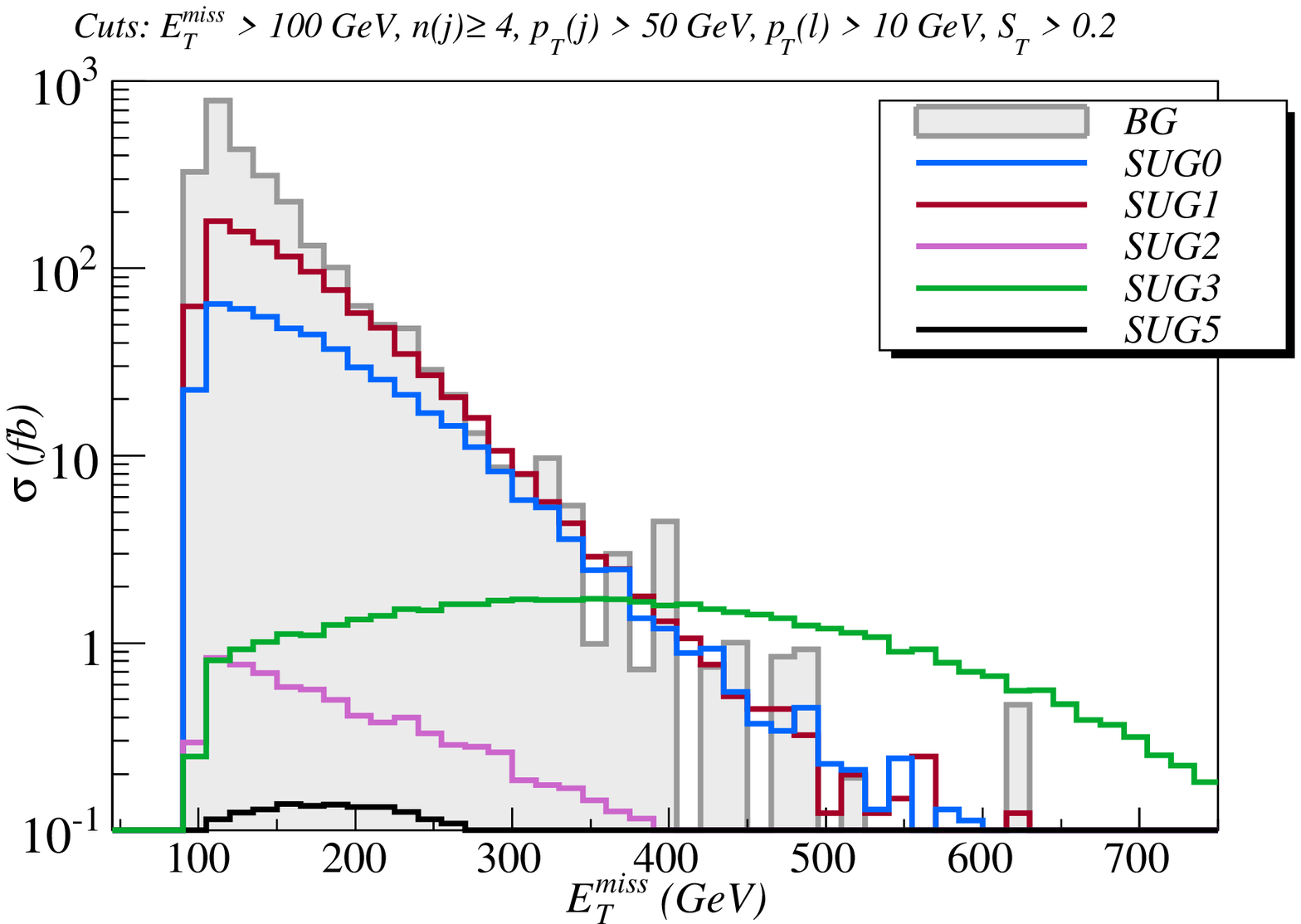}   
(b)\includegraphics[width=0.46\textwidth]{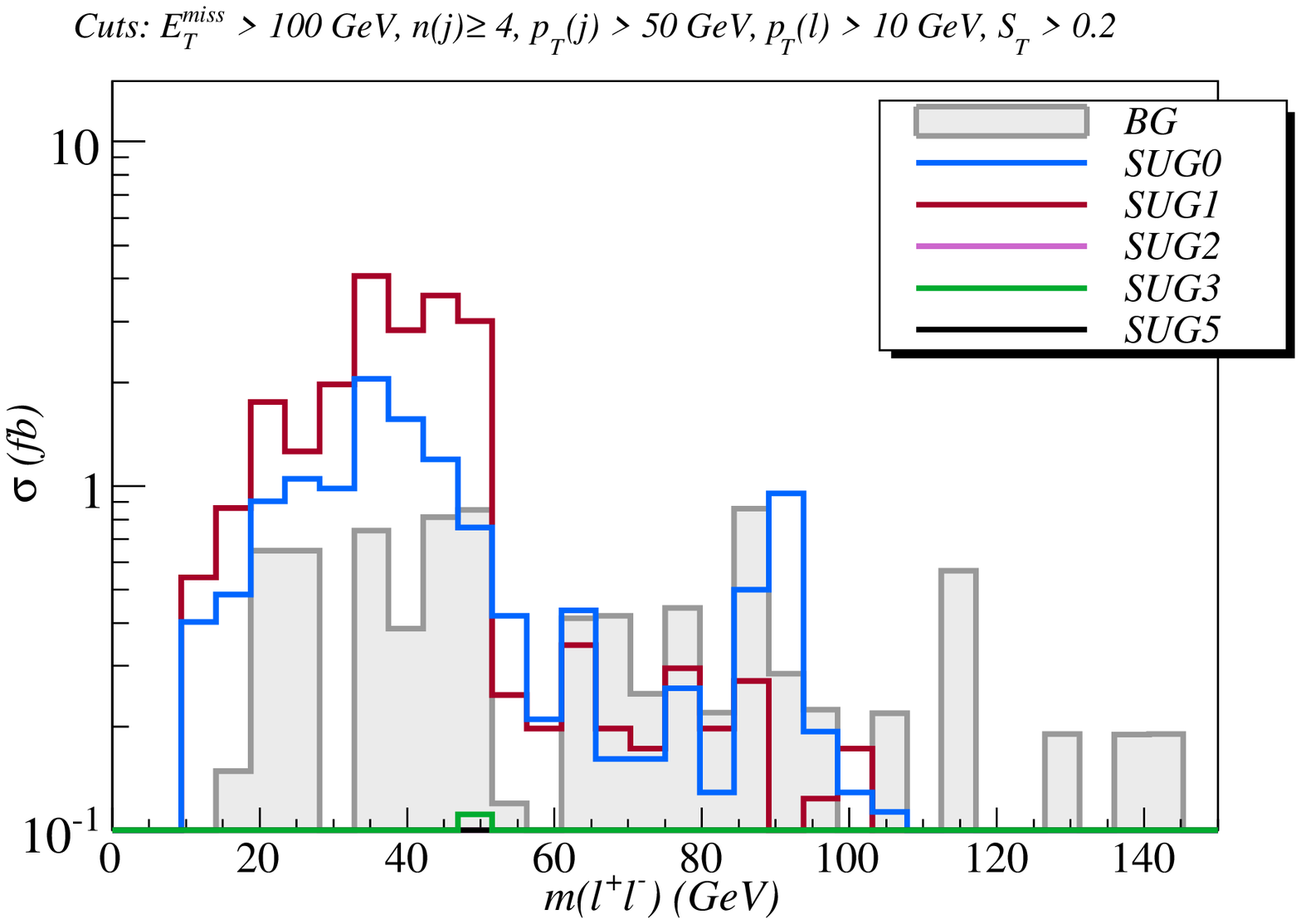}\\[2mm]
(c)\includegraphics[width=0.46\textwidth]{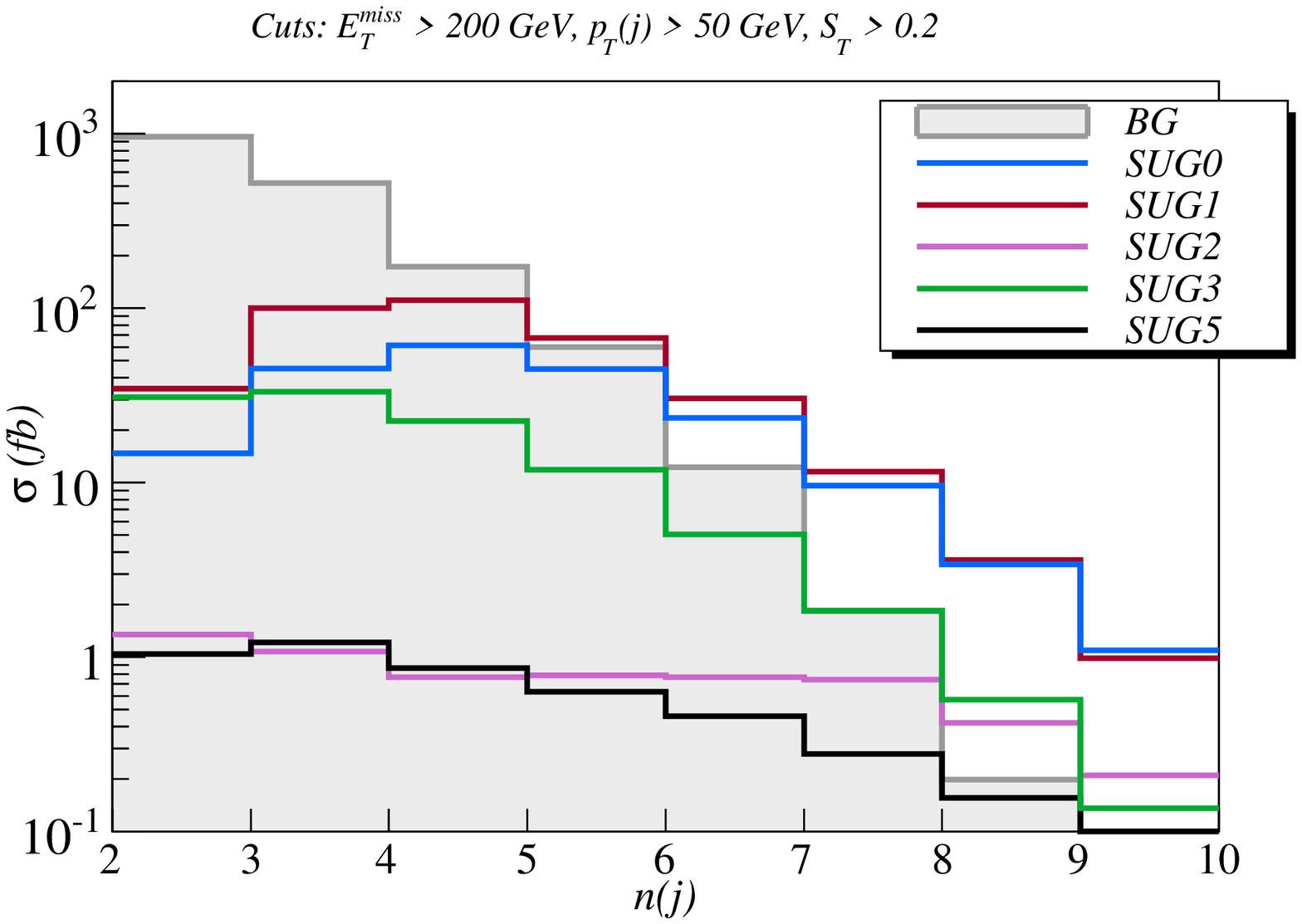}
\caption{{\protect\small $E_T^{miss}$, $m_{l^+l^-}$ and $n(j)$ distributions calculated for the LHC at 7 TeV CM energy. In frames (a) and (b) the cuts imposed are $E_{T}^{miss}>\;100\;{\rm GeV},\;n_{jets}\ge 4,\;p_{T}(j)>50\;{\rm GeV},\;p_T(l)>10\;{\rm GeV}\;{\rm and}\;S_{T}>0.2$. In frame (c) the cuts imposed are $E_{T}^{miss}>\;200\;{\rm GeV},\;n_{jets}\ge 2,\;p_{T}(j)>50\;{\rm GeV},\;{\rm and}\;S_{T}>0.2$. The blue, red, purple, green and black curves correspond to the SUG$i$ points listed in Table~\ref{tab:bm}. The SM background is shaded in grey. 
 }}
\label{dists0}
\end{figure}

In order to probe the discovery potential of the LHC at $\sqrt{s} = $7 TeV and with 1 fb$^{-1}$ of integrated luminosity we first consider the following set of cuts:
\begin{itemize}
\item  $E_{T}^{miss} > 100$ GeV, $n(j) \ge 4$, $p_T(j) > 50$ GeV, $p_T(l) \ge 10$ GeV and $S_T > 0.2$
\end{itemize}
where $S_T$ is the transverse sphericity. 
Figures~\ref{dists0}(a) and (b) show the $E_{T}^{miss}$ and opposite sign/same flavor dilepton invariant mass ($m_{l^+l^-}$) distributions for the five CMSSM points from Table~\ref{tab:bm} along with the SM background (BG). The BG was generated using AlpGen~\cite{alpgen} and Pythia~\cite{pythia} and includes all the processes listed 
in \cite{Baer:2010tk}. The SUSY decay branching ratios 
were computed using the SUSYHIT~\cite{susyhit} package and the signal events were generated using Pythia's SLHA interface. For more details on the Monte Carlo  simulation, see Ref.~\cite{Baer:2010tk}.

The signals from points SUG0 and SUG1 are dominated by gluino pair production with subsequent 3-body decays to quarks plus a neutralino or chargino, resulting in 
an $E_{T}^{miss}$ distribution slightly harder than that of the BG, 
as seen in Figure~\ref{dists0}(a). On the other hand, point SUG3 has a heavier gluino, but lighter squarks and its cross-section is dominated by 
squark-pair and gluino-squark production. 
In this case the lighter $\tq_R$ states decay mainly to $\tz_1 + q$, since $\tz_1$ is mostly a bino state.  Therefore the SUG3 point presents a much harder $E_{T}^{miss}$ distribution, peaking around 350 GeV. 
This is different for SUG2: despite having a larger total cross-section than point SUG3, the events from SUG2 are largely dominated by chargino/neutralino pair production, resulting in a much softer $E_T^{miss}$ spectrum. Finally, the $A^0/H^0$ funnel region, represented by point SUG5, has too small cross-sections to be seen during the first run of the LHC.

We also show the $m_{l^+l^-}$ ($l =e,\mu$) distributions for the same set of cuts. 
The ones from points SUG0 and SUG1 exhibit  
the characteristic edge at $m_{\tz_2} - m_{\tz_1}$, while the BG is mainly featureless, since the $E_{T}^{miss}$ cut vetoes most of the $Z \to l^+l^-$ events. Although the $m_{l^+l^-}$ distribution can in principle be used to extract the $\tz_2-\tz_1$ mass gap, the curves shown in Figure~\ref{dists0}(b) are at the few fb level, what makes a measurement during the first LHC run improbable. We also point out that despite having a sizeable cross-section after cuts, the $\ttau$ co-annihiliation point (SUG3) is poor in dilepton events, since it has a light stau ($m_{\ttau_1} = 164$ GeV), which enhances $\tz_2 \to \ttau_1 + \tau$ decays (40\% branching ratio)
and suppress decays to $\mu$'s and $e$'s (7\%). 

Finally, Figure~\ref{dists0}(c) shows the jet multiplicity $n(j)$ distributions 
for the points SUG0--SUG5 and the BG. Here, a harder cut of $E_{T}^{miss}>200$~GeV 
is used, while the $p_{T}$ requirements for jets and leptons and the $S_T$ cut are
the same as above. As one can see, at SUG0 and SUG1 the signal exceeds the BG 
for $n(j)\ge 6$ and $n(j)\ge 5$, respectively, while at SUG2 and SUG3 
this is the case only for $n(j)\ge 8$.

To determine the best search strategy for our reference points, we next optimize over a grid of cuts including number of jets, $E_T^{miss}$, number of leptons, number of b-jets and $p_{T}$s of first and second jets, following the same procedure as  outlined in Ref.~\cite{Baer:2010tk}. The discovery channel is defined as the set of cuts which satisfies $S > max[5,5\sigma,0.2B]$ and maximizes $S/\sqrt{S+B}$, where $S$ ($B$) is the number of signal (background) events. 
The ATLAS collaboration estimates~\cite{atlastdr} that the systematic 
uncertainty on the data-driven background determination can range from 
20\% to 50\% for 1~fb$^{-1}$. In order to incorporate these uncertainties in our discovery analysis, we take the conservative value of 50\% for the total background systematical uncertainty and add it in quadrature to the statistical error ($\sqrt{B}$). The total uncertainty on the BG is then given by $\sqrt{(0.5B)^2 + B}$ and it is used to compute the signal significance $\sigma$~\cite{atlastdr}.

\begin{figure}[t]
\centering
(a)\includegraphics[width=0.46\textwidth]{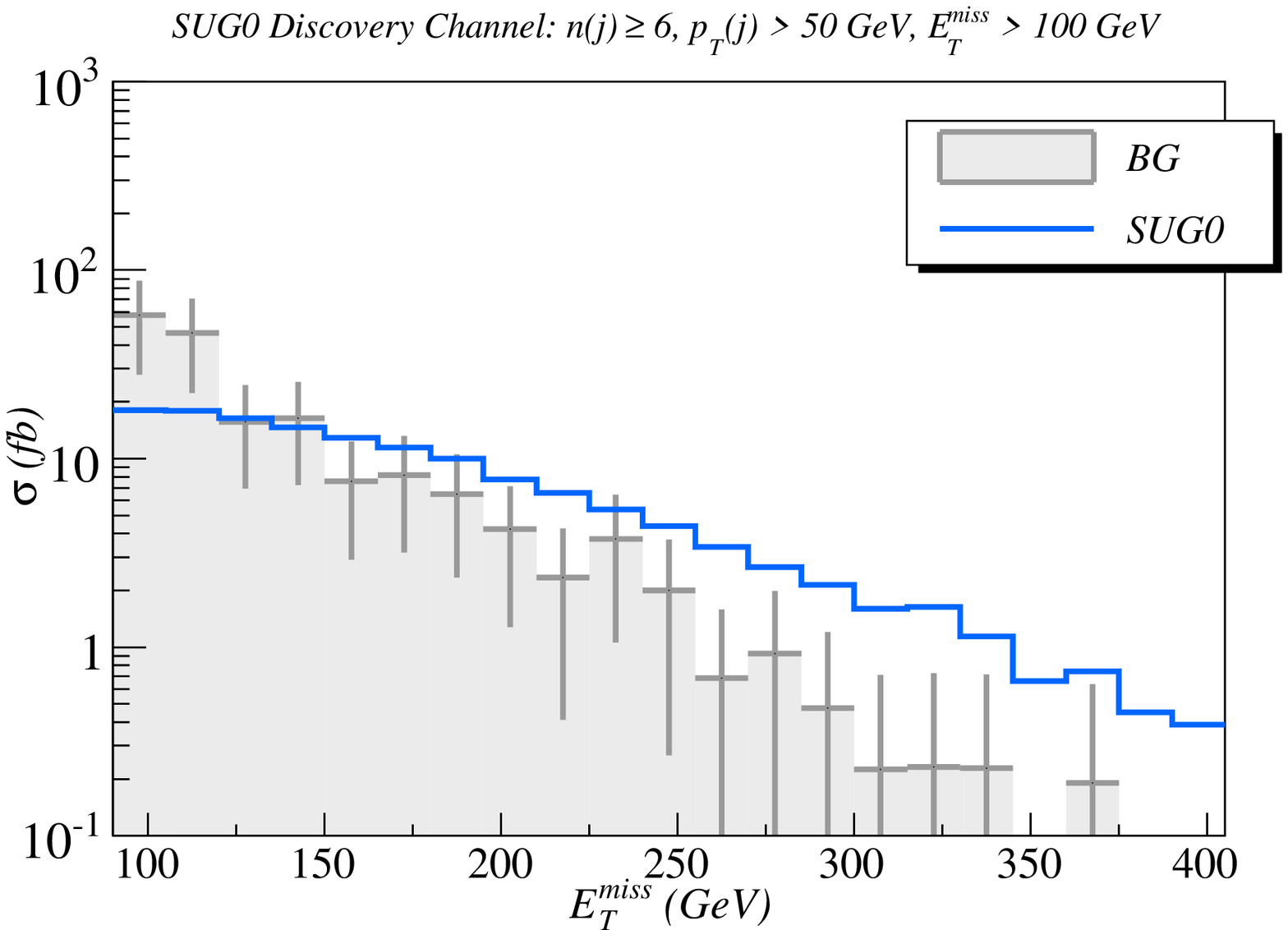}
(b)\includegraphics[width=0.46\textwidth]{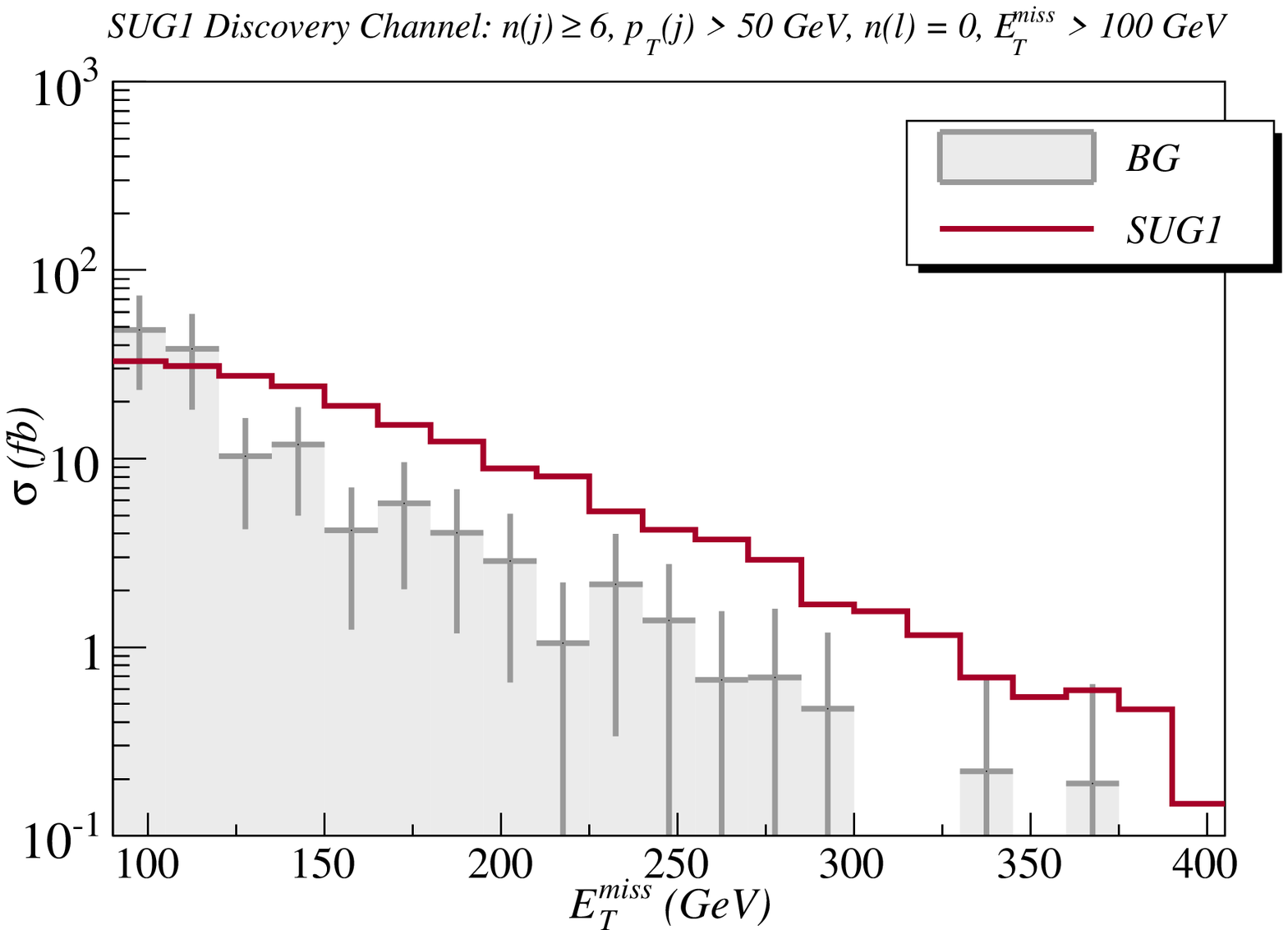}\\[2mm]
(c)\includegraphics[width=0.46\textwidth]{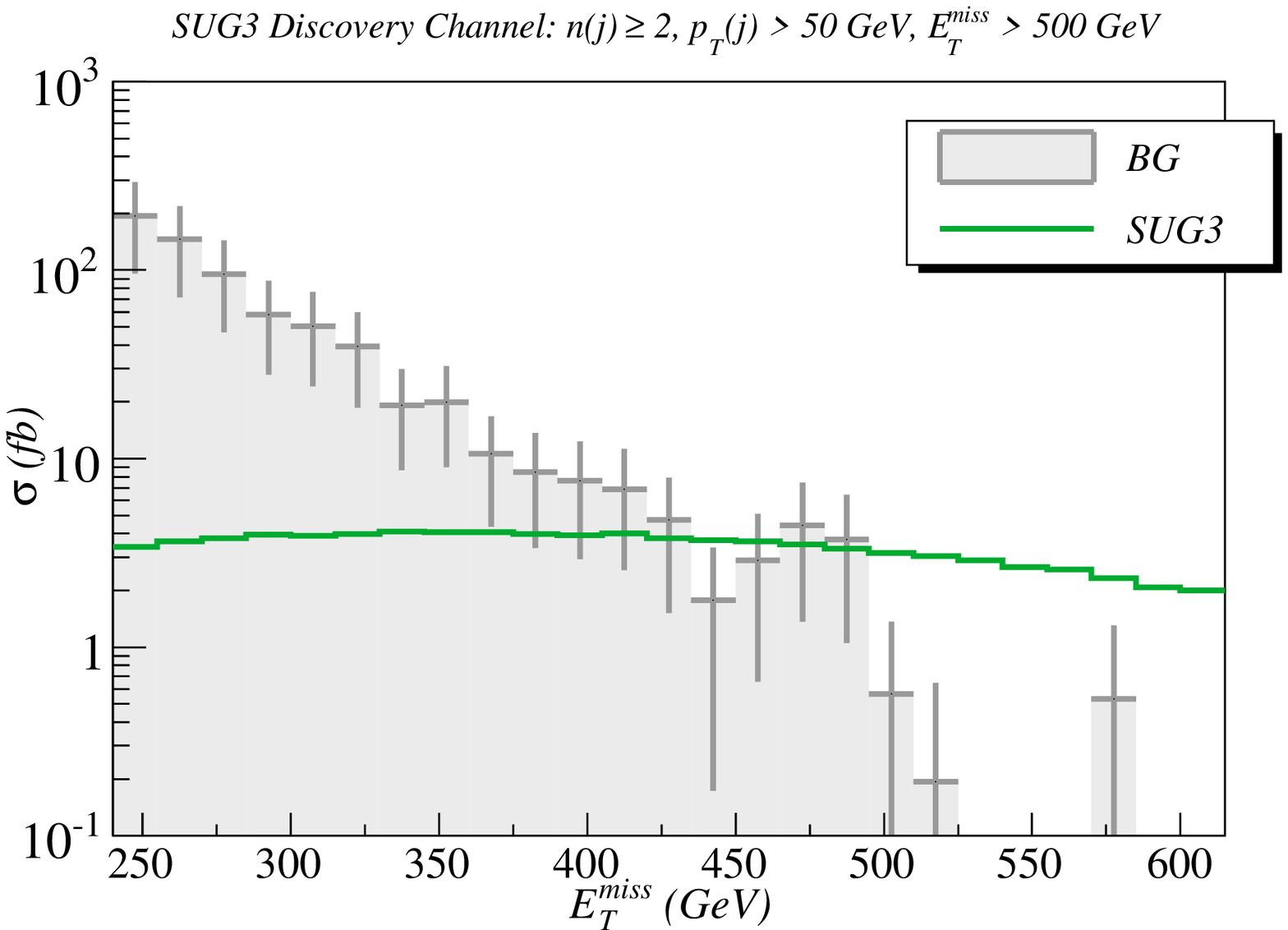}
\def\baselinestretch{1.}
\caption{{\protect\small $E_T^{miss}$ distributions for the points SUG0, SUG1 and SUG3 after the respective optimized cuts have been applied, with exception of the  $E_T^{miss}$ cut. The discovery channel (optimized set of cuts) for each point is shown on top 
of each plot. The SM background is shaded grey and the error bars correspond to the combined systematical (50\%) and statistical errors, as described in the text.}}
\label{dsc7}
\end{figure}

In Figure~\ref{dsc7} we show the $E_T^{miss}$ distribution for the points SUG0, SUG1 and SUG3 in their respective discovery channels, 
without cutting on $E_T^{miss}$. 
As mentioned above, the lowest fine-tuning point (SUG0) and point SUG1 have heavy squarks and a light gluino. The latter decays to two quarks and a neutralino/chargino through 3-body decays, resulting in a signal rich in jets. As a consequence, the optimized cuts for both points corresponds to a soft $E_T^{miss}$ cut ($> 100$ GeV), with a high jet multiplicity ($\geq 6$). However, for the SUG0 point the optimal channel is for $n(l)\ge 0$, while the SUG1 point has a better $S/\sqrt{S+B}$ ratio in the $n(l) = 0$ channel. Assuming 1 fb$^{-1}$, this optimized set of cuts gives 130 and 138 events for the SUG0 signal and background, respectively. For the SUG1 point the number of signal and background events are 182 and 103. As seen from Figures~\ref{dsc7}(a) and (b), both points are above the BG for 
$E_T^{miss} > 150$~GeV. However, both points are below the $5\sigma$ level if we assume 50\% systematical uncertainties for the BG. If the BG uncertainties can be reduced below 30\% (10\%), the point SUG1 (SUG0) would be above the discovery threshold during the first LHC run. We also point out that we have only considered simple counting signals and once evidence for a signal has been seen, the distribution shapes can be used to improve the reach potential. 
Moreover, the analysis may be further optimized by means of reference 
priors \cite{Demortier:2010sn}. 
We therefore conclude that points SUG0 and SUG1 will likely be visible during the first LHC run.

As seen from Figure~\ref{dists0}(a), the point SUG3 has a much harder $E_T^{miss}$ spectrum and a smaller production cross-section. The discovery channel in this case corresponds to a very hard $E_T^{miss}$ cut in the $n(j)\geq2$ channel. The signal is expected to have 32 events in this channel, while the SM expectation is 5 events. 
Furthermore, the SUG3 signal distribution has a very distinct shape, what can help to distinguish it from the background. We therefore conclude that the SUG3 point, 
which is representative for the $\ttau$ co-annihilation region, should  
also be visible during the first LHC run.

However, even after applying our optimization procedure, 
the signals of points SUG2 and SUG5 are not visible with 7~TeV CM energy and 
1~fb$^{-1}$ of data. We estimate that $\sim 50$ (20)~fb$^{-1}$ would be necessary for a $5\sigma$ discovery of point SUG2 (SUG5) at 7 TeV.

As shown in Figures~\ref{dists0} and \ref{dsc7}, the first LHC run at 7 TeV with 1 fb$^{-1}$ luminosity will clearly be able to find unambiguous evidence for non Standard Model physics  scenarios with a light gluino or sub-TeV squarks, in agreement with the results of Refs. \cite{Baer:2010tk} and \cite{nelson2010}. However, once a signal is seen, it will be important to distinguish between the different regions of SUSY parameter space and indeed to check whether it is a gluino and/or squarks that are initially produced. The accessible scenarios for the first LHC run, represented by the points SUG0, SUG1 and SUG3, have distinct properties that can provide an indication of which class of events is realized.  
Here we focus on three main distinguishing characteristics: 
the $E_T^{miss}$, $n(j)$ and $m_{l^+l^-}$ distributions.

From Figure~\ref{dists0}(a) we see that the amplitude of the signal of 
each point is directly related to its respective gluino mass. 
Therefore the observed number of signal events could point to the gluino mass scale. 
Region~1 (red) has a light LSP in the 50--60~GeV mass range, 
see Fig.~\ref{dmcdmsboundft100}, 
in order to have a $h^0$ resonant annihilation. Consequently the gluino 
is also light, with a mass between 400 and 500~GeV 
(assuming $\Delta < 100$). On the other hand, the $\ttau$ co-annihilation region 
(region 3) features heavier LSPs of 120--250~GeV mass and a 
small $\ttau_1$--$\tz_1$ mass difference; this results in 
$800~{\rm GeV}<m_{\tilde g}<1400$~GeV, as seen in Figure~\ref{higgsp1}(b). Therefore, these two scenarios, represented by points SUG1 and SUG3, belong to disconnected regions in the CMSSM parameter space. Finally, the remaining low-fine-tuned region that does not satisfy the WMAP bound, represented by the SUG0 point, lies in the intermediate gluino mass range, with $450~{\rm GeV}<m_{\tilde g}<750$~GeV. If a signal is observed during the first LHC run, the number of observed events can indicate the gluino mass scale. 
Unless $m_{\tilde g}$ lies in the overlap region, which is 
disfavoured by dark matter considerations, 
it could point to the relevant scenario.

Another important discriminant between the $\ttau$ co-annihilation and $h^0$ 
resonant annihilation regions is the jet multiplicity. Since the former has 
$m_{\tg} \sim m_{\tq}$, in most cases the jets from gluino decays will be soft 
and escape detection. Therefore the signal will mostly consist of 2--4 jet 
events coming from squark cascade decays. For the red points (region 1), 
on the other hand, 
the light gluinos will go through 3-body decays and generate events rich in jets. 
As a consequence, region 1 will most likely be primarily observed in the multijet 
channel, while region 3 (green points) will be visible at lower jet multiplicities, 
as already indicated by the discovery channels discussed above. 
Figure~\ref{dists0}(c) shows the jet multiplicities for $E_T^{miss} > 200$ GeV 
and $p_T(j) > 50$ GeV. Although the SUG3 point is not visible in this channel, 
it illustrates the behavior just described: the SUG3 signal peaks at low jet 
multiplicities ($n(j) = 2$) and points SUG1 and SUG0 show a peak at $n(j) = 4$. 

So far the $E_{T}^{miss}$ and $n(j)$ distributions seem to provide useful tools to distinguish between regions 1 and 3.  However, both  
these distributions are similar for the SUG0 and SUG1 points and give little hope of distinction between these two scenarios. On the other hand,  
the dilepton invariant-mass distribution in Figure~\ref{dists0}(b) shows an interesting distinction between points SUG0 and SUG1. While both points display the $m_{\tz_2}-m_{\tz_1}$ mass edge, only the events from SUG0 have a visible peak at the $Z$ mass. Such a peak is characteristic of the low-fine-tuned region in Figure~\ref{higgsp} that does not satisfy the WMAP bound, in which there are significant gluino decays to the heavier neutralinos $\chi_{(3,4)}^{0}$. This is not the case for region 2 (red points), 
as here 
$\mu$ is systematically larger\footnote{For the decays of the gluino to $\chi_{(3,4)}^0$, the fraction of points where the channel is kinematically accessible is: $\sim100\%$ for the low fine-tuned region that does not satisfy the WMAP bound, $\sim 50\%$ for the light red points in Figure~\ref{higgsp} that sub-saturate the relic density and $\sim 10\%$ for the dark red points that saturate the relic density.}, as seen in Table~\ref{tab:bm}. However, as the total number of Z di-lepton events expected in the 7 TeV run is only about 3, one will have to wait for the second stage of LHC running to use this measure to distinguish between these scenarios. 
Nevertheless the $m_{l^+l^-}$ distribution can still provide another important information about the underlying model. As discussed above, the excess of dilepton events at low invariant mass is characteristic of $\tz_2 \to \tz_1 + l^+l^-$ decays with $m_{\tz_2} - m_{\tz_1} < m_Z$. For the $h^0$ resonance annihilation (red region), the mass difference of the two lightest neutralinos has a rough Gaussian distribution of $55 \pm 3$ GeV ($90\%$ of all points are within the range 49--61~GeV). Therefore, if the neutralino mass difference is greater than this, but the observed signal still is consistent with a light gluino ($m_{\tg} \lesssim 500$ GeV), then CMSSM neutralino dark matter is not likely. Finally, the observation of a signal rich in dijets 
(and taus) plus $E_T^{miss}$ but poor in $e/\mu$  
dilepton events would point to the SUG3 scenario.

%------------------------------------------------------------------------------
\subsection{LHC at 14 TeV}
%------------------------------------------------------------------------------

As shown in the previous section, the first LHC run will be able to test a considerable portion of the low fine-tuned CMSSM, the exceptions being the $A^0/H^0$ resonant annihilation (SUG5) and the low $\mu$, high $m_0$ (SUG2) regions. While the latter will be accessible to the next generation of dark matter direct detection experiments, it is still desirable to have a corroboratory signal at the LHC. Therefore we now address whether the LHC operating at its design CM energy of $\sqrt{s}=14$~TeV 
would be able to probe low-fine-tuned models in regions 2 and 5.

\begin{figure}[t]
\centering
(a)\includegraphics[width=0.46\textwidth]{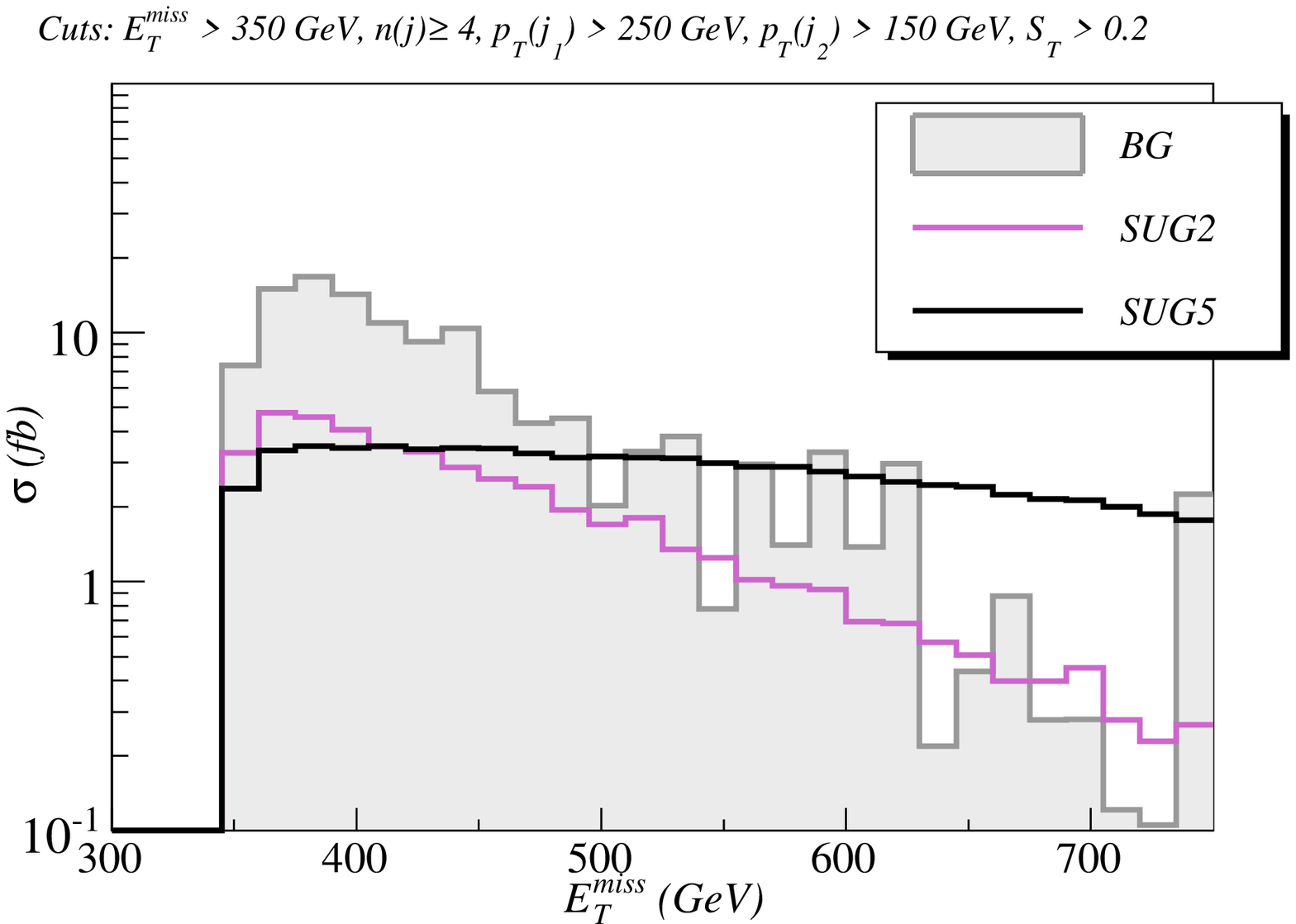}
(b)\includegraphics[width=0.46\textwidth]{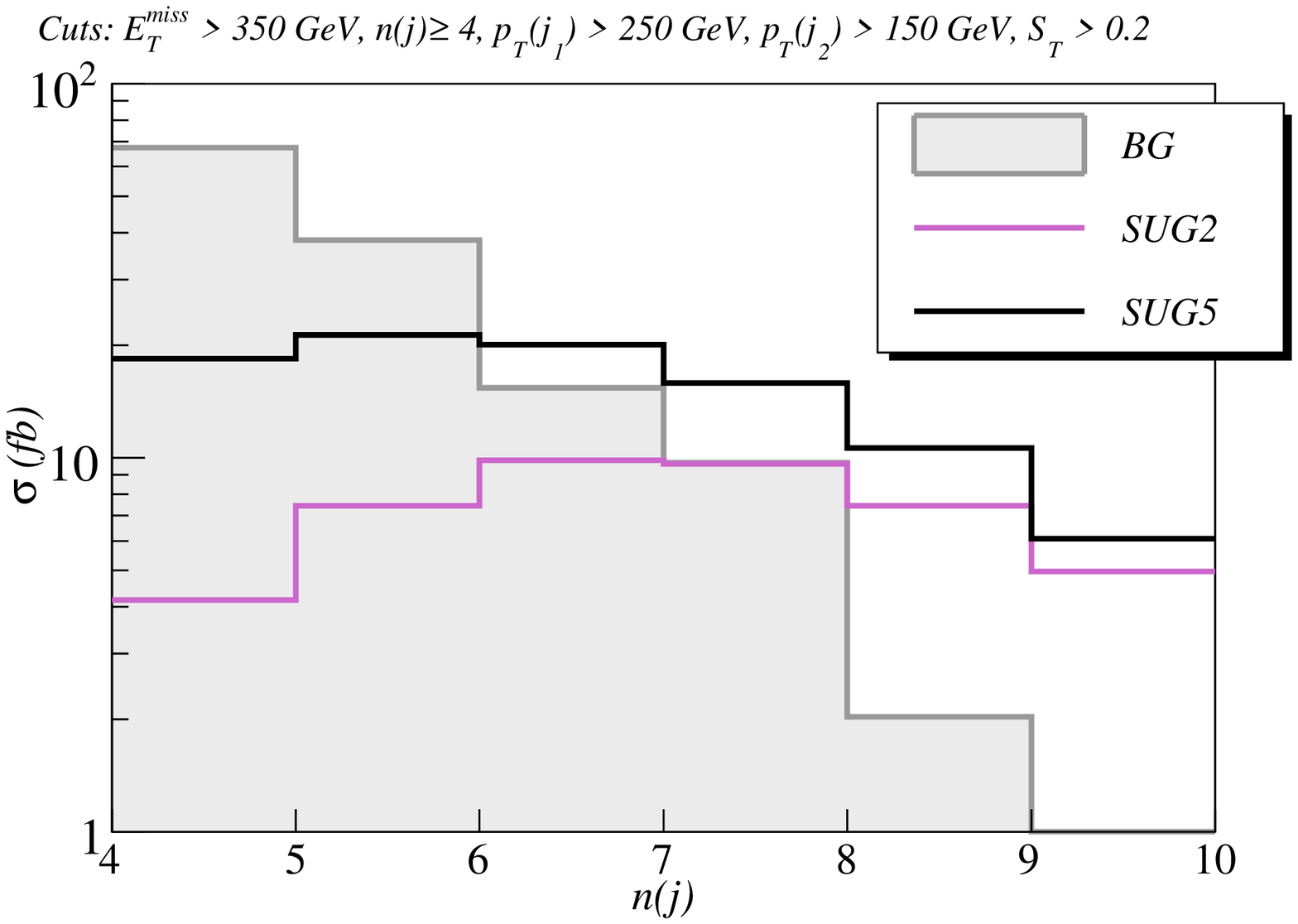}\\[2mm]
(c)\includegraphics[width=0.46\textwidth]{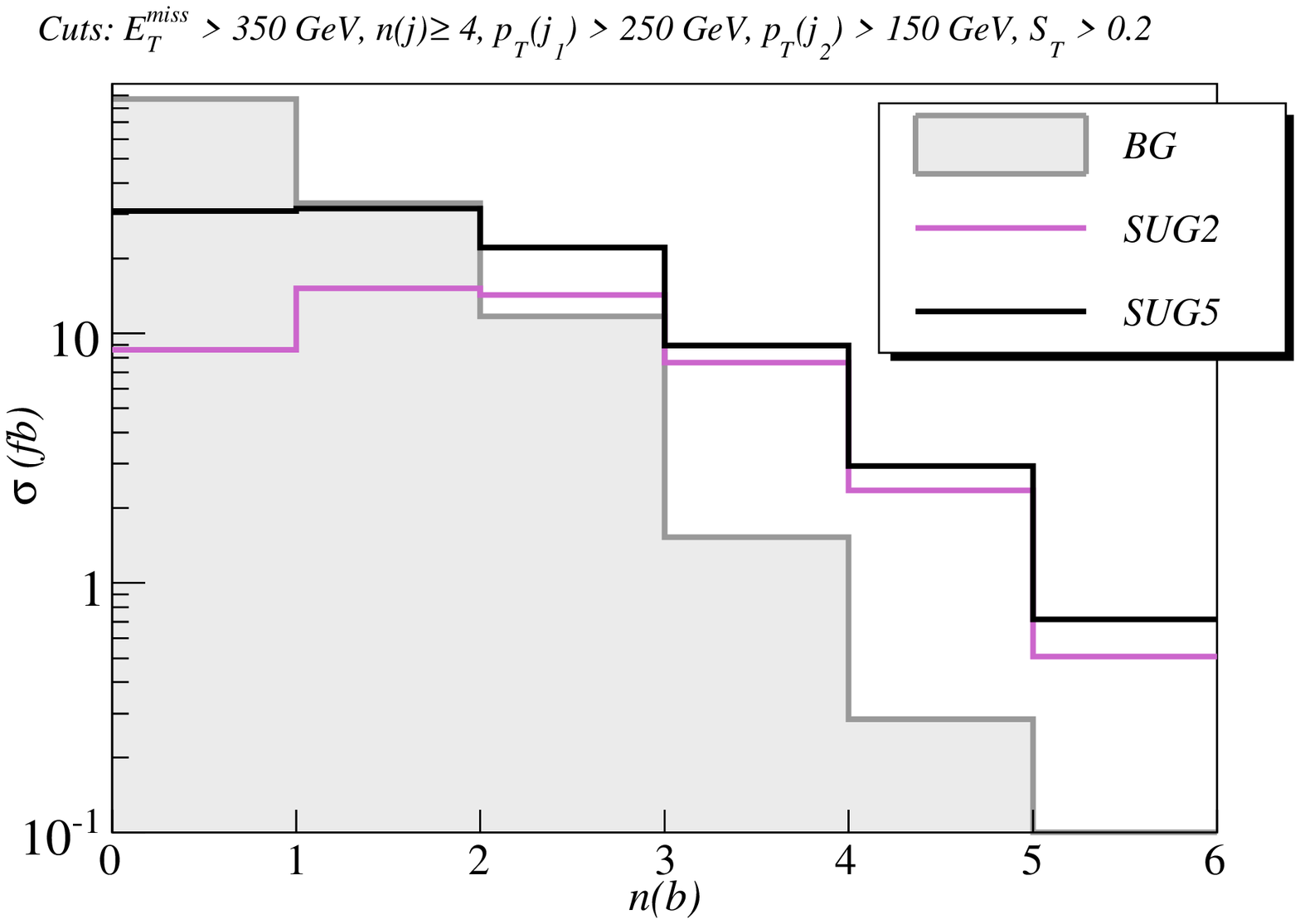}
\def\baselinestretch{1.}
\caption{{\protect\small $E_T^{miss}$, $n(j)$ and $n(b)$ distributions calculated for the LHC at $\sqrt{s}=14$~TeV. The cuts imposed are $E_{T}^{miss}>\;350\;{\rm GeV},\;n_{jets}\ge 4,\;p_{T}(j_1)>250\;{\rm GeV},\;p_{T}(j_2)>150\;{\rm GeV},\;{\rm and}\;S_{T}>0.2$. The black and purple curves correspond to the SUG5 and SUG2 points, respectively. The SM BG is shaded in grey. 
 }}
\label{dists1}
\end{figure}

To this aim, Figure~\ref{dists1} shows the distributions of 
$E_T^{miss}$, number of jets and number of $b$-jets for points SUG2 and SUG5 along with the respective SM BG distributions. The following set of cuts has been applied:
\begin{itemize}
\item  $E_{T}^{miss} > 350$ GeV, $n(j) \ge 4$, $p_T(j_1) > 250$ GeV, $p_T(j_2) > 150$ GeV and $S_T > 0.2$
\end{itemize}

The SUG2 events that pass the cuts come mostly from gluino-pair production. Since the $\tst_1$ is much lighter than the first and second generation squarks, the gluino decays $\sim 80\%$ of the time to third generation quarks. As a result, there is a large $b$-jet multiplicity in the SUG2 signal, as shown in Figure~\ref{dists1}(c). This point would hence be easily visible above BG in the $n(b) \geq 3,4$ channels. 
The SUG5 point has a heavier gluino ($m_{\tg} = 1.2$ TeV) and lighter 1st/2nd generation squarks, so the signal after cuts is dominated by gluino-squark production and the $n(b)$ distribution is softer than in the SUG2 scenario. However, due to the large branching ratio for $\tq_R \to \tz_1 + q$ decays, the SUG5 point has a hard $E_T^{miss}$ spectrum, similar (in shape) to the SUG3 point. Thus the SUG5 signal could be observed in the multijets channel with a hard $E_T^{miss}$ cut, as seen in  Figure~\ref{dists1}(b). We estimate that, for $\sqrt{s} = 14$ TeV, a $5\sigma$ discovery for both points would require an integrated luminosity of 
$\mathcal{L} \lesssim 10\; {\rm fb}^{-1}$.

\section{Conclusions}
  
In summary, using the fine-tuning measure, we have made a detailed study of the possibility of testing in the near future the CMSSM as a solution to the hierarchy problem.  Broadly, the regions of low fine-tuning split into two characteristic classes. The first class has light gluinos or light squarks and will likely be tested in the 7~TeV run at the LHC. The second class has a heavy gluino but the LSP has a significant higgsino component;   
this class is testable by direct dark matter searches in the near future. 
Together, these complementary experiments will be able to cover almost all of the parameter space with fine-tuning $\Delta<100$. 
To cover all of this parameter space by SUSY searches at the LHC will require 
running at the full 14~TeV CM energy. 
In addition, the low-fine-tuned regions can be tested 
indirectly by Higgs searches covering the mass range $m_h\le 120$~GeV.

\section{Acknowledgements}

The research presented here was partially supported by the EU ITN grant 
UNILHC 237920 (``Unification in the LHC era'') 
and by the French ANR project ToolsDMColl, BLAN07-2-194882. 
SC is supported by the UK Science and Technology Facilities Council  
under contract PPA/S/S/2006/04503. 
The work of DG is supported in part by CNRS research contract 225933 at 
Ecole Polytechnique and by the ERC advanced grant 226371 (``Mass TeV''). 
AL is partially supported by the Fulbright Program and CAPES.

%------------------------------------------------------------------------------


\begin{thebibliography}{99}
%------------------------------------------------------------------------------

\bibitem{Cassel}
S.~Cassel, D.~M.~Ghilencea and G.~G.~Ross,
  %``fine-tuning as an indication of physics beyond the MSSM,''
  Nucl.\ Phys.\  B {\bf 825} (2010) 203  [arXiv:0903.1115 [hep-ph]];
  Phys.\ Lett.\  B {\bf 687} (2010) 214  [arXiv:0911.1134 [hep-ph]];
  Nucl.\ Phys.\  B {\bf 835} (2010) 110  [arXiv:1001.3884 [hep-ph]].
 
\bibitem{Allanach:2001kg} 
B.~C.~Allanach, 
  %``SOFTSUSY: A C++ program for calculating supersymmetric spectra,'' 
  Comput.\ Phys.\ Commun.\ \textbf{143}
  (2002) 305 [arXiv:hep-ph/0104145].

\bibitem{Ellis:1986yg}
J.~R.~Ellis, K.~Enqvist, D.~V.~Nanopoulos and F.~Zwirner,
  %``Observables In Low-Energy Superstring Models,''
  Mod.\ Phys.\ Lett.\  A {\bf 1} (1986) 57; 
  %%CITATION = MPLAE,A1,57;%
%\bibitem{Barbieri:1987fn} 
R.~Barbieri and G.~F.~Giudice, 
  %``Upper Bounds On Supersymmetric Particle Masses,''
  Nucl.\ Phys.\ B \textbf{306} (1988) 63. 
  %%CITATION = NUPHA,B306,63;%% 
 
%\cite{Degrassi:2002fi}
\bibitem{Degrassi:2002fi}
  G.~Degrassi, S.~Heinemeyer, W.~Hollik, P.~Slavich and G.~Weiglein,
  %``Towards high-precision predictions for the MSSM Higgs sector,''
  Eur.\ Phys.\ J.\  C {\bf 28} (2003) 133  [arXiv:hep-ph/0212020].
  %%CITATION = EPHJA,C28,133;%%

\bibitem{Feng:2000bp} 
J.~L.~Feng and K.~T.~Matchev, 
  %``Focus Point Supersymmetry: Proton Decay, Flavor and 
  %CP Violation, and the Higgs Boson,'' 
  Phys.\ Rev.\ D \textbf{63} (2001) 095003 [arXiv:hep-ph/0011356]; 
  %%CITATION = PHRVA,D63,095003;%%
J.~L.~Feng, K.~T.~Matchev and T.~Moroi,
  %``Multi-TeV scalars are natural in minimal supergravity,'' 
  Phys.\ Rev.\ Lett.\ \textbf{84} (2000) 2322
  [arXiv:hep-ph/9908309]; 
K.~L.~Chan, U.~Chattopadhyay and P.~Nath,
  % ``Naturalness, weak scale supersymmetry and the prospect for 
  % the observation of supersymmetry at the Tevatron and at the LHC,'' 
  Phys.\ Rev.\ D \textbf{58} (1998) 096004
  [arXiv:hep-ph/9710473]. 
  

\bibitem{Gogoladze:2009bd}
\revision{
  I.~Gogoladze, M.~U.~Rehman and Q.~Shafi,
  %``Amelioration of Little Hierarchy Problem in SU(4)(c) x SU(2)(L) x
  %SU(2)(R),''
  Phys.\ Rev.\  D {\bf 80} (2009) 105002
  [arXiv:0907.0728 [hep-ph]].
  %%CITATION = PHRVA,D80,105002;%%
  }

\bibitem{Horton:2009ed}
D.~Horton and G.~G.~Ross,
  %``Naturalness and Focus Points with Non-Universal Gaugino Masses,''
  Nucl.\ Phys.\  B {\bf 830} (2010) 221
  [arXiv:0908.0857 [hep-ph]].
  
\bibitem{Belanger:2006is}
G.~Belanger, F.~Boudjema, A.~Pukhov and A.~Semenov,
  %``micrOMEGAs: Version 1.3,''
  Comput.\ Phys.\ Commun.\  {\bf 174} (2006) 577 [arXiv:hep-ph/0405253];
  %%CITATION = CPHCB,174,577;%%
  %``micrOMEGAs2.0: A program to calculate the relic density of dark matter 
  %in a generic model,''
  Comput.\ Phys.\ Commun.\  {\bf 176} (2007) 367 [arXiv:hep-ph/0607059].
  %%CITATION = CPHCB,176,367;%%

\bibitem{Dunkley:2008ie}
  J.~Dunkley {\it et al.}  [WMAP Collaboration],
  %``Five-Year Wilkinson Microwave Anisotropy Probe (WMAP) Observations:
  %Likelihoods and Parameters from the WMAP data,''
  Astrophys.\ J.\ Suppl.\  {\bf 180} (2009) 306
  [arXiv:0803.0586 [astro-ph]].
  %%CITATION = APJSA,180,306;%%
  
\bibitem{Baer:2010tk}
  H.~Baer, V.~Barger, A.~Lessa and X.~Tata,
  %``Capability of LHC to discover supersymmetry with \sqrt{s}=7 TeV and 1 fb^{-1},''
  JHEP {\bf 1006} (2010) 102 [arXiv:1004.3594 [hep-ph]];
  %H.~Baer, V.~Barger, A.~Lessa and X.~Tata,
  %``Supersymmetry discovery potential of the LHC at $\sqrt{s}=$10 and 14 TeV
  %without and with missing $E_T$,''
  JHEP {\bf 0909} (2009) 063 [arXiv:0907.1922 [hep-ph]].
  
\bibitem{cms}
CMS~Collaboration,
  %``Search for Supersymmetry in pp Collisions at 7 TeV in Events with Jets and
  %Missing Transverse Energy,''
  arXiv:1101.1628 [hep-ex].
  %%CITATION = ARXIV:1101.1628;%%
  
\bibitem{daCosta:2011qk}
\revision{
  J.~B.~G.~da Costa {\it et al.}  [Atlas Collaboration],
  %``Search for squarks and gluinos using final states with jets and missing
  %transverse momentum with the ATLAS detector in sqrt(s) = 7 TeV proton-proton
  %collisions,''
  arXiv:1102.5290 [hep-ex].
  %%CITATION = ARXIV:1102.5290;%%
}

\bibitem{Ahmed:2009zw}
  Z.~Ahmed {\it et al.}  [The CDMS-II Collaboration],
  %``Dark Matter Search Results From The Cdms II Experiment,''
  Science {\bf 327} (2010) 1619
  [arXiv:0912.3592 [astro-ph.CO]].
  %%CITATION = SCIEA,327,1619;%%

\bibitem{cassel-thesis}
S.~Cassel, University of Oxford D.Phil thesis, to appear.
 
\bibitem{Buchmuller:2007ui}
W.~Buchmuller, L.~Covi, K.~Hamaguchi, A.~Ibarra and T.~Yanagida,
  %``Gravitino dark matter in R-parity breaking vacua,''
  JHEP {\bf 0703} (2007) 037 [arXiv:hep-ph/0702184]; 
  %%CITATION = JHEPA,0703,037;%%
%\bibitem{Covi:2009bk}
L.~Covi, J.~Hasenkamp, S.~Pokorski and J.~Roberts,
  %``Gravitino Dark Matter and general neutralino NLSP,''
  JHEP {\bf 0911} (2009) 003 [arXiv:0908.3399 [hep-ph]].
  %%CITATION = JHEPA,0911,003;%%

\bibitem{Baer:2010gr}
H.~Baer, S.~Kraml, A.~Lessa and S.~Sekmen,
  %``Thermal leptogenesis and the gravitino problem in the Asaka-Yanagida
  %axion/axino dark matter scenario,''
  arXiv:1012.3769 [hep-ph] and references therein.
  %%CITATION = ARXIV:1012.3769;%%

%\cite{Covi2}
\bibitem{Covi2}
  L.~Covi, H.~-B.~Kim, J.~E.~Kim, L.~Roszkowski,
  %``Axinos as dark matter,''
  JHEP {\bf 0105 } (2001)  033.
  [hep-ph/0101009].

\bibitem{alpgen} 
M. Mangano, M. Moretti, F. Piccinini, R. Pittau and A. Polosa,
  JHEP {\bf 0307} (2003) 001 
  [arXiv:0206293 [hep-ph]].

\bibitem{pythia} 
T. Sjostrand, S. Mrenna and P. Skands,
  JHEP {\bf 0605} (2006) 026 
  [arXiv:0603175 [hep-ph]].

\bibitem{susyhit}
A.~Djouadi, M.~M.~Muhlleitner and M.~Spira,
  %``Decays of Supersymmetric Particles: the program SUSY-HIT
  %(SUspect-SdecaY-Hdecay-InTerface),''
  Acta Phys.\ Polon.\  B {\bf 38} (2007) 635
  [arXiv:hep-ph/0609292].
  %%CITATION = APPOA,B38,635;%%

\bibitem{atlastdr}
G.~Aad {\it et al.} (Atlas Collaboration)
     arXiv:0901.0512 [hep-ex];
     arXiv:1003:3124 [hep-ex]. 

\bibitem{Demortier:2010sn}
L.~Demortier, S.~Jain and H.~B.~Prosper,
  %``Reference priors for high energy physics,''
  Phys.\ Rev.\  D {\bf 82} (2010) 034002
  [arXiv:1002.1111 [stat.AP]]; 
  %%CITATION = PHRVA,D82,034002;%%
S.~Sekmen, talk at SUSY10, 23--28 Aug.\ 2010, Physikalisches Institut, Bonn, Germany.

\bibitem{nelson2010}
B.~Altunkaynak, M.~Holmes, P.~Nath, B.~Nelson and G.~Peim,
  Phys.\ Rev.\ D \textbf{82} (2010) 115001
  [arXiv:1008.3423 [hep-ph]].
  %%CITATION = 1008.3423;%%

\end{thebibliography}
\end{document}